# Ultra-fast perpendicular Spin Orbit Torque MRAM

Murat Cubukcu*,[1], Olivier Boulle[1, †], Nikolaï Mikuszeit[1], Claire Hamelin[1], Thomas Brächer[1], Nathalie Lamard[6], Marie-Claire Cyrille[6], Liliana Buda-Prejbeanu[1], Kevin Garello[2], Ioan Mihai Miron[1], O. Klein[1], G. de Loubens[4], V. V. Naletov[1,4,5], Juergen Langer[3], Berthold Ocker[3], Pietro Gambardella[2] and Gilles Gaudin[1]

1 Univ. Grenoble Alpes, CEA, CNRS, Grenoble INP, INAC, SPINTEC, F-38000 Grenoble, France

[2] Department of Materials, ETH Zurich, Hönggerbergring 64, CH-8093 Zürich, Switzerland

[3] Singulus Technologies, Hanauer Landstr, 103, 63796, Kahl am Main, Germany

[4] Service de Physique de l'Etat Condensé (CNRS URA 2464), CEA Saclay, 91191 Gif-sur-Yvette, France

[5] Institute of Physics, Kazan Federal University, Kazan 420008, Russian Federation

[6] CEA Leti, F-38000 , Grenoble, France

**We demonstrate ultra-fast (down to 400 ps) bipolar magnetization switching of a three-terminal perpendicular Ta/FeCoB/MgO/FeCoB magnetic tunnel junction. The critical current density rises significantly as the current pulse shortens below 10 ns, which translates into a minimum in the write energy in the ns range. Our results show that SOT-MRAM allows fast**

**and low-power write operations, which renders it -promising for non-volatile cache memory applications.**

The introduction of non-volatility at the cache level is a major challenge to the IT industry as it would lead to a large decrease of the power consumption of microprocessors by minimizing their static and dynamic power consumption and pave the way towards normally-off/instant-on computing. Among other technologies, STT-MRAM has been identified as a promising candidate for the non-volatile replacement of SRAM cache memory technology[1]. STT-MRAM combines CMOS compatibility, high retention time (10 years), large endurance and relatively fast write time (down to 4 ns for reliable switching in perpendicular STT-MRAM[2]). However, cache memory applications typically require faster operations (~ns for L1 cache) combined with a large endurance due to their high access rate. Very fast switching (sub-ns) has been recently demonstrated using stacks where the magnetization~~s~~ of the free and the fixed layer~~s~~ are perpendicular[3]–[5]. However, this gain in operation speed comes at the expense of a rise in the current flowing through the tunnel barrier. As a consequence, manufacturers are currently facing reliability issues due to the accelerated aging of the tunnel barrier when injecting these high write current densities[6], [7]. Another drawback of STT-MRAM is that reading and writing use the same current path. This results in an undesired writing during the read-out of the bit[7] as well as a high read power since the tunnel barrier needs to have a very small resistance to sustain the large writing current densities.

Recently, we have proposed a novel memory concept, named Spin-Orbit Torque-MRAM (SOT-MRAM), that combines the STT advantages and naturally solves the above mentioned issues[8]–[10]. The memory is based on the discovery that a current flowing in the plane of a magnetic multilayer with structural inversion asymmetry, such as Pt/Co/$AlO_x$, exerts a torque on the magnetization, which can lead to magnetization reversal[9], [11], [12]. Such a torque arises from the conversion of the orbital to spin angular momentum through the spin Hall effect in the heavy metal and/or the Rashba-Edelstein effect at the interfaces[9], [13]–[15]. The key advantage of the SOT-MRAM is that writing and reading are decoupled due to their independent current paths. Thus, the SOT-MRAM intrinsically solves the reliability issues in current STT-MRAM promising a potentially unlimited endurance.

To be a strong candidate for non-volatile cache memory applications, SOT-MRAM needs to be fast. We recently demonstrated deterministic switching induced by current pulses shorter than 200 ps in dots made of Pt/Co/AlOx stacks[12]. However, the use of a Pt seed layer in MgO based MTJ does not allow to reach the high TMR ratio needed for memory applications[16], [17] (>100%), as it promotes a (111) fcc texture while a (100) bcc structure at the CoFe/MgO interface is needed to achieve high TMR [18]–[20]On the contrary, the Ta/FeCoB/MgO/FeCoB MTJ stacks commonly used for STT-MRAM seem ideal for SOT-MRAM since they combine a high TMR, a perpendicular magnetization[21] and a large spin Hall effect in Ta[22]. In this article, we demonstrate that magnetization switching can be achieved by very short current pulses (down to 400 ps) in Ta/FeCoB/MgO three-terminal SOT-MRAM memory cells. Our results show that SOT-MRAM allows for fast, and low-power write operations, rendering it promising for non-volatile cache memory applications.

The magnetic tunnel junctions (MTJ) was deposited by magnetron sputtering using a Singulus Timaris® deposition machine with the following structure[23]–[25] 10 Ta/1 $Fe_{60}Co_{20}B_{20}$/MgO/ 1.3 $Fe_{60}Co_{20}B_{20}$/0.3 Ta/$FM_1$/Ru0.85/$FM_2$ (thicknesses in nm), where $FM_1$=[0.4 Co/ 0.4 Cu/ 1.4 Pt]$_{x5}$ /0.6 Co and $FM_2$=0.6 Co/[0.4 Cu/ 1.4 Pt/ 0.4 Co]$_{x12}$/ 0.4 Cu/2 Pt (see Fig. 1(a)). Functional three-terminal single cells with lateral dimensions down to 150 nm diameter on top of a 330 nm wide Ta track were fabricated as described in Ref.[26] The results presented here are obtained from a sample with a 275 nm diameter MTJ on top of a 635 nm Ta track (see Fig.1 (b)). All measurement are carried out at room temperature. Figure 1 (c) shows a typical TMR hysteresis cycle corresponding to the successive reversal of the FeCoB (1 nm) free layer and pinned layer, the magnetic field being applied perpendicularly to the sample plane. A TMR of up to 55%, associated with a sharp reversal of the magnetization of the free layer, is observed. The resistance area product of the junction is about 600 $\Omega.\mu m^2$. For the current induced magnetization switching experiments, current pulses are injected in the Ta bottom track using a fast voltage pulse generator whereas the TMR signal is measured using a DC voltage source connected to the MTJ in series with a 1M$\Omega$ resistor. This resistor prevents high voltages spikes on the MTJ during the pulse injection. A 100 $\Omega$ resistor was connected in parallel to the track to minimize the impedance mismatch. The pulse rise time is 220 ps for pulse widths $\tau_P$ <2 ns, and 1.5 ns for wider pulses. The pulse width is defined as the full width at half maximum.

Figure 2 (a) shows the TMR signal measured after the pulse injection as a function of the amplitude of the current pulse injected in the track. An in-plane magnetic field $\mu_0 H_{ip}$=100 mT is applied along the current direction to allow for the bipolar switching[9]. The current pulse is 550 ps long. Starting from the low resistance state and increasing the current, a sharp increase in the TMR signal is observed above a positive threshold pulse amplitude, demonstrating the reversal of the magnetization of the FeCoB bottom free layer from the parallel (P) to the anti-parallel (AP) configurations of the magnetizations. From the AP configuration, a large enough negative current allows to go back to the P configuration. This demonstrates the writing of a perpendicular SOT-MRAM memory cell by a 550 ps current pulse and its reading by the TMR signal. Note that the switching current for the P to AP switching is slightly lower than the one for the AP to P, which can be explained by the dipolar interaction between the bottom free layer and the not fully compensated synthetic antiferromagnet. The corresponding switching-current density is about $3.3x10^{12}$ A/m$^2$. The switching probability from the P to the AP configuration as a function of the-amplitude of the current for different pulse widths is plotted in Fig. 2(b) (each point is an average over 30 events) [27]. Interestingly, recent time-resolved X-ray microscopy imaging of the spin orbit torque driven magnetization reversal of Pt/Co/AlOx dots revealed that the switching probability measured electrically is not the probability of an on/off event, but more likely the fraction of the magnetic layer area that has switched[28]. The same measurements show that there are no ringing or after – pulse effects associated to switching.

Magnetization switching is observed in the whole range of pulse widths from 400 ps to 2.5 µs and is bipolar: positive currents lead to a magnetization switching from P to AP, whereas negative currents lead to a switching from AP to P. The switching current $I_c$ strongly depends on the pulse length $\tau_P$ (see Fig. 3(a)). For $\tau_P$>10 ns, $J_c$ changes little with $\tau_P$ and scales approximately linearly on log($\tau_P$) suggesting a thermally activated regime where stochastic fluctuations help the magnetization to overcome the reversal energy barrier [29], [30]. For $\tau_P$< 10 ns, a large increase of $J_c$ is observed as $\tau_P$ decreases and $J_c$ scales linearly on 1/$\tau_P$ (see Fig.3(a) inset). For $\tau_P$< 10 ns, a large increase of $J_c$ is observed as $\tau_P$ decreases and $J_c$ scales linearly on 1/$\tau_P$ (see Fig.3(a) inset). This scaling is reminiscent of early spin transfer torque predictions [31] and experiments [32] where the injected spin current polarization is aligned along the uniaxial anisotropy axis. Such a scaling is expected from the conservation of spin angular momentum, assuming the magnetization is spatially homogeneous (macrospin approximation). A different scaling is however expected in our spin orbit torque geometry where the current spin polarization is aligned perpendicular to the uniaxial anisotropy axis [12], [33]. On the other hand, several experimental study have shown that for lateral sizes typically larger than 50 nm, the magnetization reversal by spin transfer and spin orbit torques occurs by domain nucleation followed by domain wall propagation[28], [34]–[41]. In such a case, a $1/\tau_P$ scaling of the critical current is expected, which expresses that the switching time is the time for a nucleated domain wall to travel across the dot.

As expected, the switching current depends also on the external in-plane magnetic field $H_{ip}$ and decreases as $H_{ip}$ increases (Fig. 3(a)). The corresponding write energy $E=RI^2\tau_P$ is plotted in Fig. 3(b) as function of $\tau_P$, assuming it is dissipated in a 3 kΩ resistance standing for the Ta track and the addressing transistor. The energy depends non-monotonously on $\tau_P$ with a large increase of the energy as $\tau_P$ decreases for $\tau_P$<1 ns. Interestingly, a minimum in the write energy is observed between 1 and 3 ns. This feature is explained by the crossover between the thermally activated regime for large pulse width and the short pulse width regime. The energy scale extrapolated for a 50 nm wide and 3 nm thick Ta track is shown in blue on the right vertical axis A write energy of about 95 fJ at 1.5 ns can be reached, associated with a write current of about 180 µA, which is similar to the best results obtained so far for current perpendicular STT-MRAM technology[42], [43].

In conclusion, we demonstrate ultra-fast (down to 400 ps) bipolar and deterministic writing of perpendicular three-terminal spin-orbit torque (SOT)-MRAM single cells with a Ta/CoFeB/MgO/CoFeB MTJ structure. The switching current density rises significantly as the pulse shortens below 10 ns. This translates into a write energy minimum in the ns range.

These experimental results extrapolate to a switching current of around 180 µA at 1.5 ns for 50 nm track width. This makes SOT-MRAM promising for a power efficient non-volatile cache memory application.

This work was supported by the European Commission under the Seventh Framework Program (Grant Agreement 318144, spOt project), the French Agence Nationale de la Recherche (ANR) through Project SOSPIN and the Swiss National Science Foundation (Grant No. 200021-153404). V.V.N. acknowledges support from programs Competitive Growth of KFU and CMIRA'Pro of the region Rhone-Alpes. The devices were fabricated at the Plateforme de Technologie Amont (PTA) in Grenoble.

* Current address : Cavendish laboratory, University of Cambridge, JJ Thomson Avenue, Cambridge, CB3 0HE UK,

† e-mail: olivier.boulle@cea.fr

## Figure caption

Fig.1 (a) Sketch of the three-terminal MTJ. (b) Scanning electron microscopy image of a 275 nm diameter MTJ on top of a 635 nm wide Ta track. (c) Resistance as a function of the magnetic field applied perpendicularly to the sample plane.

Fig. 2 (a) TMR as a function of the current pulse amplitude $I_P$ ($\tau_P$=0.55 ns long) in the presence of an external in-plane magnetic field $\mu_0 H_{ip}$=100 mT. The TMR is measured after the injection of the current pulse. The arrows show the sweep direction of $I_P$. (b) Switching probability ($P_{sw}$) from the P to the AP configuration as a function of $I_P$ for three different pulse lengths $\tau_P$=0.55 ns (black, square), $\tau_P$=0.89 ns (red, circles) and $\tau_P$=1 ns (blue, circles) at an applied field $\mu_0 H_{ip}$=100 mT.

Fig.3 (a) Switching current $I_c$ as a function of the current pulse length $\tau_P$ for two values of the external in-plane magnetic field (P to AP switching). Inset: $I_c$ vs $1/\tau_P$ for $\mu_0 H_{ip}$ = 100 mT. (b) Energy dissipated in a 3 kΩ resistor (simulating the resistance of the Ta track and the transistor) as a function of $\tau_P$ for two values of $H_{IP}$ using the write current for the three-terminal device with a 635 nm wide Ta track. The blue scale on the right shows the write energy extrapolated for a 50 nm wide and 3 nm thick Ta track.

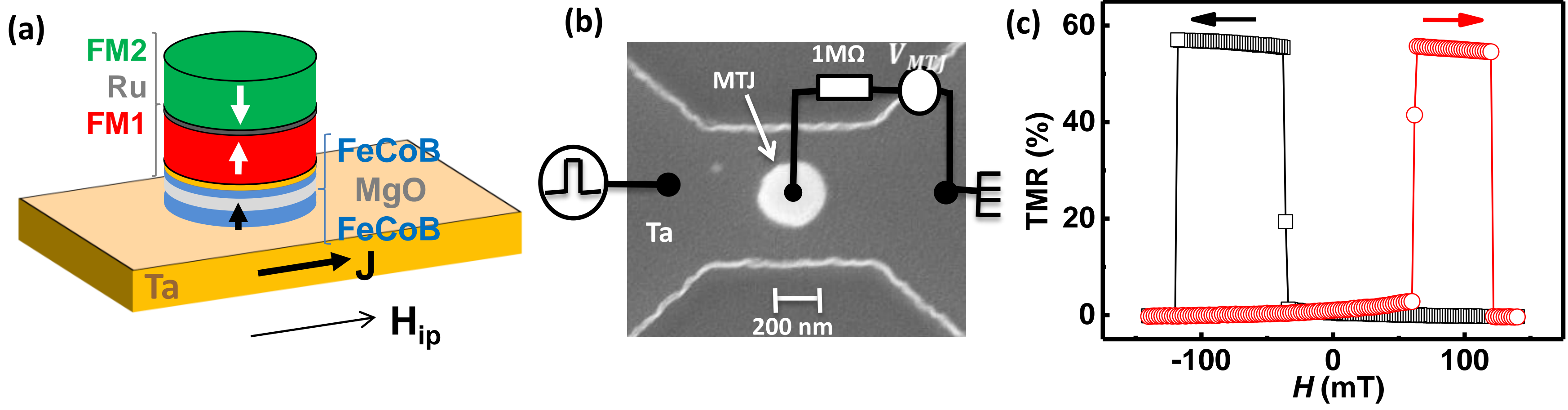

**Fig. 1 Cubukcu *et al.***

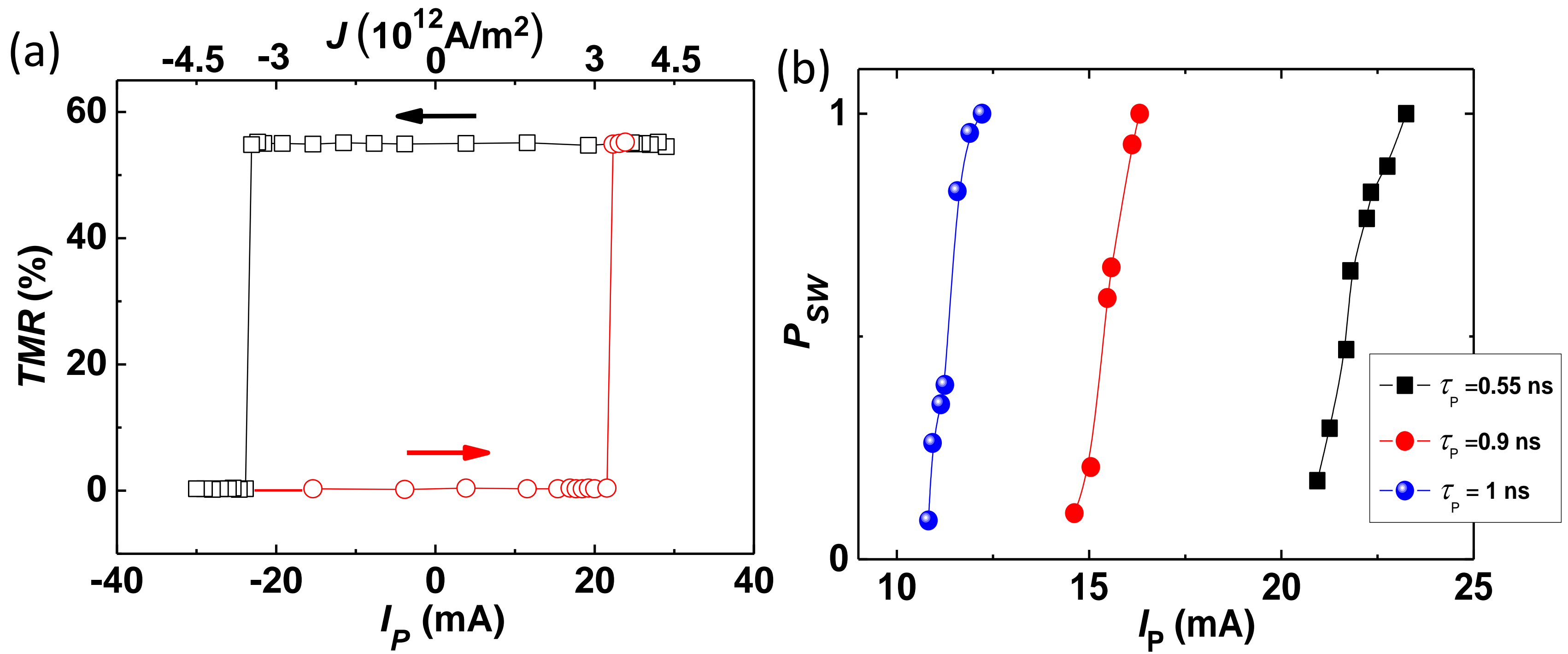

**Fig. 2. Cubukcu *et al***

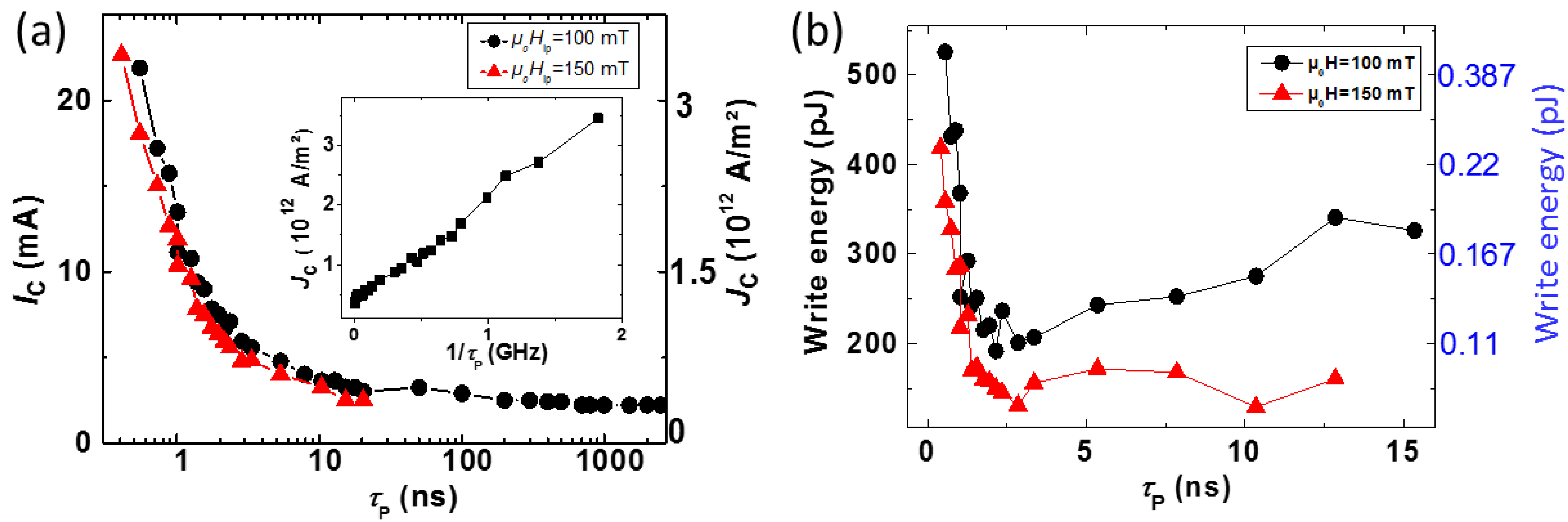

**Fig. 3. Cubukcu *et al.***